\documentclass{elsart}

\usepackage{amssymb}
\usepackage{amsfonts}
\usepackage{amsmath}
\usepackage{verbatim}
\usepackage[dvipdf]{graphicx}
\usepackage{subfigure}

\def\Tr{\textrm}
\def\ra{\rangle}            
\def\la{\langle}            
\def\dd{\textrm{d}}
\def\Bf{\boldsymbol}

\def\rv{\Bf{r}}

\def\pv{\Bf{p}}

\begin{document}

\begin{frontmatter}



\title{Low energy theory of a single vortex and electronic quasiparticles\\ in a $d$-wave superconductor}


\author{Predrag Nikoli\'c\corauthref{cor1}}
\author{and Subir Sachdev}
\corauth[cor1]{Corresponding author. email:
pnikolic@fas.harvard.edu}
\address{Department of Physics, Harvard University, Cambridge MA 02138}

\begin{abstract}
We highlight the properties of a simple model (contained in our
recent work) of the quantum dynamics of a single point vortex
interacting with the nodal fermionic quasiparticles of a $d$-wave
superconductor. We describe the renormalization of the vortex motion
by the quasiparticles: at $T=0$, the quasiparticles renormalize the
vortex mass and introduce only a weak sub-Ohmic damping. Ohmic (or
`Bardeen-Stephen' damping) appears at $T>0$, with the damping
co-efficient vanishing $\sim T^2$ with a universal prefactor.
Conversely, quantum fluctuations of the vortex renormalize the
quasiparticle spectrum. A point vortex oscillating in a harmonic
pinning potential has no zero-bias peak in the electronic local
density of states (LDOS), but has small satellite features at an
energy determined by the pinning potential. These are proposed as
the origin of sub-gap LDOS peaks observed in scanning tunneling
microscopic studies of the LDOS near a vortex.
\end{abstract}

\end{frontmatter}

\section{The Model}

A more complete discussion of the quantum dynamics of a vortex and
nodal quasiparticles in a two-dimensional $d$-wave superconductor
appears in our recent papers \cite{DWVorAct,LorenzLDOS,DiracLDOS}
which also describe the connections to microscopic theory. Here, we
focus on a simple low-energy model, which captures all the essential
features. Our model has no zero-bias peak in the electronic LDOS at
the vortex center (see also \cite{Tesh2}), in contrast to
computations in the traditional BCS framework
\cite{Wang95,machida,Franz98}. The zero-point quantum motion of the
vortex in the pinning potential leads to sub-gap satellite peaks in
the LDOS. Finally, as discussed in earlier work \cite{CompOrd1} (and
not reviewed here) a proper accounting of Aharanov-Bohm-like phases
in the vortex motion leads to periodic spatial modulations in the
LDOS (see also \cite{Tesh1}). These features have the potential to
explain the key characteristics of scanning tunneling microscopic
(STM) observations of vortices in the cuprate superconductors
\cite{aprile,hoffman,fischer}.

The degrees of freedom of our model are ({\em i\/}) the low energy
$S=1/2$ fermionic quasiparticles in the vicinity of the nodes of a
$d$-wave superconductor, described by the second-quantized Nambu
spinor $\Psi (\rv)$, and ({\em ii\/}) a (point) vortex, described by
its first-quantized position $\rv_\Tr{v}$ and canonically conjugate
momentum $\pv_\Tr{v}$. In the absence of vortex pinning (considered
later), the Hamiltonian has the very simple form ($\hbar=1$):
\begin{equation}\label{Model}
H = \sum_{\Tr{nodes}}
    \int \dd^2 r \Psi^{\dagger}(\rv) H_{D}(\rv, \rv_\Tr{v}) \Psi(\rv) \ ,
\end{equation}
where $H_D$ is the Dirac Hamiltonian in the presence of a gauge
field $\Bf{a}$ \cite{Tesh2}:
\begin{equation}\label{BdG}
H_{D} = v_{\textrm{f}} \left( \frac{\partial}{i\partial x} + a_x
(\rv,\rv_\Tr{v}) \right) \sigma^z +
      v_{\Delta} \left( \frac{\partial}{i\partial y} + a_y
(\rv,\rv_\Tr{v}) \right) \sigma^x \ .
\end{equation}
Here $\hat{\Bf{x}}$ denotes any of the four nodal directions in
momentum space and $\hat{\Bf{y}}$ the direction perpendicular to it,
$\sigma^{x,y,z}$ are Pauli matrices, $v_{\Tr{f}}$ and $v_{\Delta}$
are Fermi and gap velocities respectively, and the gauge field
$\Bf{a}$ is given by
\begin{equation}\label{VGF}
\Bf{a}(\rv,\rv_\Tr{v}) =
  \frac{\hat{\Bf{z}} \times (\rv - \rv_\Tr{v})}{2 |\rv - \rv_\Tr{v}|^2} \ ,
\end{equation}
so that it corresponds to a $\pi$-flux centered at the instantaneous
vortex position $\rv_\Tr{v}$; in the Lagrangian formulation,
Eq.~(\ref{VGF}) would appear as a Chern-Simons term. An assumption
is that the vortex core size can be neglected in the cuprates,
especially since there are no localized quasiparticle states in the
$d$-wave vortex cores. The Hamiltonian ~(\ref{BdG}) is derived from
the full Bogoliubov-de Gennes Hamiltonian in the limit of an extreme
type-II superconductor by application of the Franz-Te\v sanovi\'c
unitary transformation \cite{Tesh2}.

We have not displayed the `Doppler shift' terms in $H_{D}$: their
influence is considered elsewhere \cite{DWVorAct}, and they do not
qualitatively modify the results below. Note also that a bare vortex
Hamiltonian has not been explicitly included: the needed terms are
generated by integrating out the quasiparticles.

\section{Vortex dynamics in $d$-wave superconductors}
\label{sec:vortex}

First, we consider the influence of the nodal quasiparticles on the
vortex dynamics. This has been previously studied in a semiclassical
theory \cite{volovik,kopnin}. Our results differ from this
semiclassical theory which, we suspect, does not properly capture
the ``quantum critical'' aspects of the Dirac fermion dynamics.

Instead of specifying a vortex Hamiltonian $H_\Tr{v}$, we formulate
an imaginary-time path-integral for quasiparticles coupled to a
single vortex at the variable position $\Bf{r}_\Tr{v}(\tau)$, and
integrate out the fermionic quasiparticle fields. The result is an
effective vortex action $S_\Tr{v}$ that contains only the
quasiparticle contribution to vortex dynamics. We express $S_\Tr{v}$
in the frequency domain as an expansion in terms of the small vortex
displacement from a fictitious origin, and calculate only the terms
up to the quadratic order. Then, on general symmetry grounds we can
write:
\begin{equation}\label{VActLT}
S_{\Tr{v}} = \frac{1}{\beta} \sum_{\omega} \Bigl\lbrack
  F_{\parallel}(\omega) |\Bf{r}_{\Tr{v}}(\omega)|^2 +
  F_{\perp}(\omega) i\hat{\Bf{z}}(\Bf{r}_{\Tr{v}}^*(\omega) \times
    (\Bf{r}_{\Tr{v}}(\omega)) \Bigr\rbrack \ ,
\end{equation}
where $F_{\parallel}(\omega)$ captures the ``longitudinal''
dynamics, such as vortex friction and inertia, while
$F_{\perp}(\omega)$ captures the ``transversal'' dynamics, such as
the quasiparticle contributions to the Magnus force.

The idea behind expanding the vortex action in terms of small vortex
displacements is to make use of the solutions for the quasiparticle
eigenstates $\psi_n$ and eigenvalues $\epsilon_n$ in presence of a
\emph{static} vortex. Various virtual transitions between these
extended quasiparticle eigenstates, caused by the vortex motion, is
what gives rise to renormalization of the parameters that
characterize vortex dynamics. Thus, the contribution of
quasiparticles to the vortex action at finite temperatures
$T=1/\beta$ is found to have the following form:
\begin{eqnarray}\label{VActFull}
S_{\Tr{v}} & = & \frac{1}{2} \sum_{n,n'} \frac{1}{\beta}
\sum_{\omega}
  \left( f(\epsilon_n) - f(\epsilon_{n'}) \right)
  \frac{i\omega(\epsilon_n - \epsilon_{n'})}{\epsilon_n-\epsilon_{n'}-i\omega}
  \nonumber \\
& & ~~~~~~~~~~~~~~~~~ \times | \Bf{r}_{\Tr{v}}(\omega) \Bf{U}_{n,n'}
|^2 \ ,
\end{eqnarray}
where $f(\epsilon)$ is the Fermi-Dirac distribution function. The
main problem is to calculate the transition matrix elements
$\Bf{U}_{n,n'} = \la \psi_n | \Bf{\nabla} | \psi_{n'} \ra$, and
carry out the summations over the quantum numbers $n$ and $n'$.
Using the microscopic model given by ~(\ref{Model}) and
~(\ref{BdG}), we find the quasiparticle eigenstates in presence of a
static vortex, and substitute them into ~(\ref{VActFull}). In
comparison with ~(\ref{VActLT}) we obtain at small frequencies:
\begin{equation}
F_{\parallel}(\omega) = -\eta |\omega| + A_1 \omega^2 \ln(|\omega|)
+
  \frac{m_\Tr{v} \omega^2}{2} + A_2 |\omega|^3 \ .
\end{equation}
 At zero temperature the nodal quasiparticle contribution to the
 vortex mass is of the order of an electron mass:
\begin{equation}
m_{\Tr{v}} \approx 0.05 \left( \frac{1}{v_\Tr{f}^2} +
           \frac{1}{v_{\Delta}^2} \right) \Lambda \ ,
\end{equation}
where $\Lambda$ is the high energy cutoff of the Dirac Hamiltonian,
and there is no vortex friction ($\eta=0$, $A_1=0$) apart from a
universal sub-Ohmic damping ($A_2(v_{\Delta}/v_{\Tr{f}}) \neq 0$).
These key results can be understood by a simple scaling argument
\cite{DWVorAct}---non-analytic terms can only arise from infrared
singular terms, and by power-counting, the first infrared
singularity arises only at order $\omega^3$. Ohmic friction arises
either at finite temperatures:
\begin{eqnarray}
\eta & = & \frac{\pi}{6} \left( \frac{1}{v_{\Tr{f}}^2} +
\frac{1}{v_{\Delta}^2} \right)
           T^2 \\
A_1 & = & -\frac{\ln(2)}{4} \left( \frac{1}{v_{\Tr{f}}^2} +
\frac{1}{v_{\Delta}^2}
           \right) T \ln(T) \ , \nonumber
\end{eqnarray}
or in presence of perturbations that create a finite density of
states at zero energy $\rho(0) \neq 0$ (the scale $T$ above is
replaced by a scale $\propto \rho(0)$, which could be, for example,
the strength of disorder, or Zeeman splitting). In general,
quasiparticles do not contribute transversal vortex dynamics
($F_{\perp}=0$), unless $\rho(0) \neq 0$. When the Doppler shift is
included in calculations, these conclusions remain qualitatively the
same in terms of their dependence on $T$, $\rho(0)$ and
$v_{\Delta}/v_\Tr{f}$, but the numerical co-efficients change
significantly.

Our findings are different from the semiclassical results in two
important ways. First, we find a finite renormalization of the
vortex mass due to quasiparticles, while the semiclassical approach
predicts a mass that diverges in small magnetic fields as
$H^{-1/2}$. Second, we find that quasiparticles cannot give rise to
vortex friction despite their gapless spectrum, unless they are
helped by a finite temperature or perturbations such as disorder. In
contrast, the semiclassical approach predicts not only that friction
is possible in the limit of infinite quasiparticle scattering time,
but also that quasiparticles can produce transversal forces on the
vortex. We hope that future experiments will resolve these
discrepancies.

\section{Quasiparticle spectra in the vicinity of a vortex}
\label{sec:qp}

Now we turn to the converse problem of the influence of the vortex
motion on the quasiparticle spectrum. From the results in
Section~\ref{sec:qp}, we know that the effective action for the
vortex acquires a mass $m_{\Tr{v}}$, and that effects of damping are
negligible. We are interested in the electronic spectrum near a
localized vortex, induced either by the repulsive interaction with
other vortices, or by a pinning potential associated with disorder.
In both cases, for small vortex displacement, it is reasonable to
make a harmonic approximation, and so we use the Hamiltonian $H
\rightarrow H + H_{\Tr{v}}$ where
\begin{equation}\label{VorHam}
H_{\Tr{v}} = \frac{\pv_\Tr{v}^2}{2m_\Tr{v}} + \frac{1}{2} m_\Tr{v}
\omega_\Tr{v}^2 \rv_\Tr{v}^2
\end{equation}
The effective vortex mass $m_\Tr{v}$, and the harmonic trap
frequency $\omega_\Tr{v}$ will be treated as given parameters, which
can be determined microscopically \cite{DWVorAct,VorMass}.

We will apply several simplifications that amount to removing all
sources of anisotropy in our model: we will set $v_\Tr{f} =
v_{\Delta}$, use a single vortex harmonic frequency, and completely
neglect the Doppler shift. Such simplifications are, of course, not
quantitatively justified in realistic circumstances, but allow
gaining a deeper physical insight and nevertheless produce
quasiparticle spectra that are remarkably similar to the ones
observed in experiments.

\subsection{Perturbation theory}

We define the operators $b^{\dagger}_\mu$ and $b_\mu$ ($\mu \in
\lbrace x,y \rbrace$ that raise and lower respectively the quantum
numbers $n_\mu$ of the vortex in the harmonic trap. By inserting the
resolution of unity in terms of the trapped vortex eigenmodes
$|n_x,n_y\ra$ we can write the Hamiltonian perturbatively as $H =
H_0 + H_1 + \cdots$, where the unperturbed Hamiltonian includes
effects of the vortex zero-point quantum fluctuations:
\begin{equation}\label{H0}
H_0 =  \omega_\Tr{v} b^{\dagger}_\mu b^{\phantom{\dagger}}_\mu +
       \int \dd^2 r \Psi^{\dagger}  V_0  \Psi^{\phantom{\dagger}} \ ,
\end{equation}
while the perturbation describes resonant scattering of
quasiparticles from the fluctuating vortex:
\begin{equation}\label{H1}
H_1 = \int \dd^2 r \Psi^{\dagger} \left( V^\mu b^{\dagger}_\mu +
h.c. \right)
     \Psi^{\phantom{\dagger}} \ .
\end{equation}
Here we have introduced the following matrix elements:
\begin{eqnarray}\label{VMatrEl}
V_0(\rv) & = & \la 0,0 | H_{D}(\rv) | 0,0 \ra \nonumber \\
V^x(\rv) & = & \la 1,0 | H_{D}(\rv) | 0,0 \ra \\
V^y(\rv) & = & \la 0,1 | H_{D}(\rv) | 0,0 \ra \nonumber \ .
\end{eqnarray}
The remaining terms in the full Hamiltonian correspond to scattering
events in which the trapped vortex undergoes two or more virtual
transitions between its discrete eigenmodes; such scattering
processes are also generated in the perturbation theory, and their
physical effects can be qualitatively obtained from ~(\ref{H0}) and
~(\ref{H1}) alone.

All calculations are performed numerically in the basis that
diagonalizes ~(\ref{H0}). After all the simplifications, the
quasiparticle eigenfunctions of $H_0$ are characterized by the
following quantum numbers: ``charge'' $q = \pm 1$, angular momentum
$l \in \mathbb{Z}$, and radial wavevector $k>0$. Their spectrum
$\epsilon = qk$ is gapless at the gap nodes, and there are no
localized states in the vortex cores of pure $d$-wave
superconductors. The perturbation ~(\ref{H1}) describes coupling of
Dirac fermions to a single bosonic two-dimensional oscillator.

The quasiparticle LDOS is obtained from the quasiparticle Green's
function $G_{l_1,k_1;l_2,k_2}(\omega)$ expressed in the spinor
representation:
\begin{eqnarray}\label{MatrLDOS}
\rho(\epsilon,\rv) & = & -\frac{1}{\pi} \Tr{sign}(\epsilon) \cdot
\Tr{Im} \Biggl\lbrace
   \sum_{l_1,l_2} \int \dd k_1 \dd k_2 \\
& & \Tr{Tr} \Bigl\lbrack
  T_{l_1,k_1}^{\phantom{\dagger}}(\rv) G_{l_1,k_1;l_2,k_2}(\epsilon)
    T_{l_2,k_2}^{\dagger}(\rv)
  \Bigr\rbrack \Biggr\rbrace \ , \nonumber
\end{eqnarray}
with $T_{l,k}(\Bf{r})$ being the ``Fourier'' weight that translates
between the position and momentum representations. Perturbative
expansion of the Green's function yields an expansion of the LDOS
that satisfies the following scaling form:
\begin{equation}\label{LDOSscaling}
\rho(\epsilon,r) = \frac{\omega_\Tr{v}}{\hbar v_\Tr{f}^2}
\sum_{n=0}^{\infty}
  \alpha^{2n} F_n \left( \frac{\epsilon}{\hbar \omega_\Tr{v}},
             \frac{\epsilon r}{\hbar v_\Tr{f}} ; \alpha \right) \ .
\end{equation}
The small parameter $\alpha$ is the ratio of the perturbation energy
scale and the vortex harmonic frequency:
\begin{equation}\label{SmallParameter}
\alpha = \left( \frac{m_\Tr{v} v_\Tr{f}^2}{\hbar \omega_\Tr{v}}
\right)^\frac{1}{2}
\end{equation}

\subsection{The quasiparticle LDOS}

\begin{figure}
\includegraphics[width=1.65in]{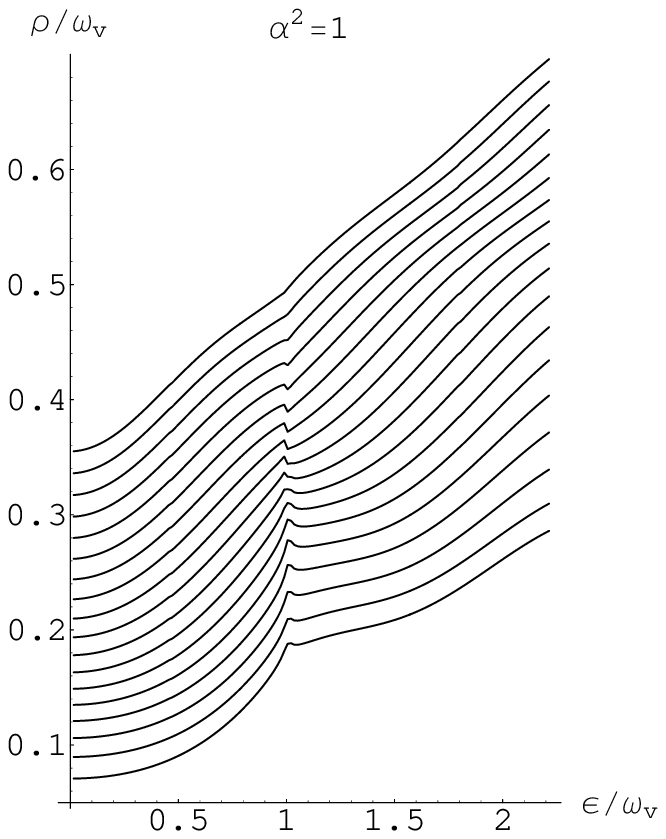}
\includegraphics[width=1.65in]{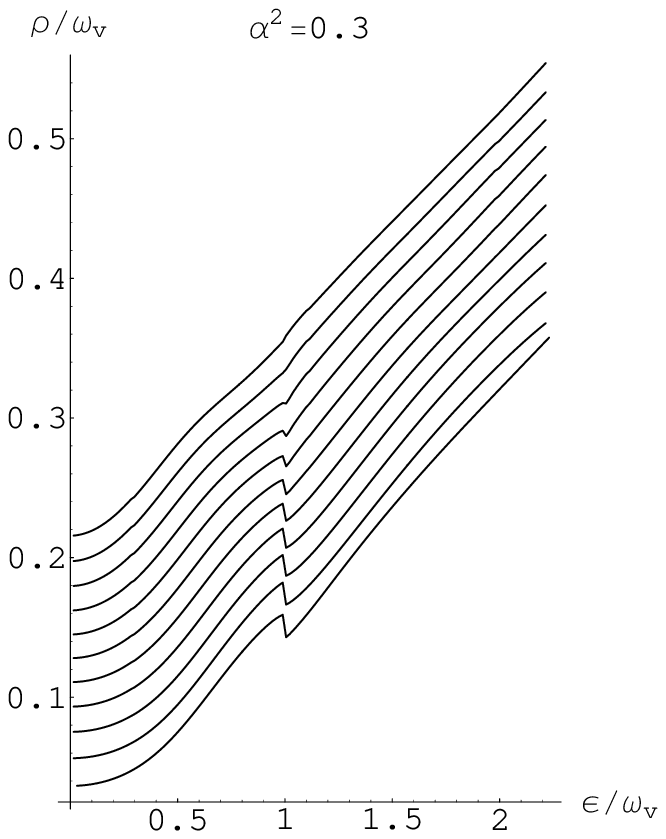}
\caption{\label{LDOStotal}Energy scans of the full LDOS at gradually
increasing distances $r$ from the vortex core. The plots are offset
vertically for clarity, starting from $r=0$ at the bottom, and
moving up with increments $\Delta r = 0.2 \omega_\Tr{v}^{-1}$ for
$\alpha^2=1$ and $\Delta r = 0.33 \omega_\Tr{v}^{-1}$ for $\alpha^2
= 0.3$ ($\Delta r \omega_\Tr{v}\approx (5\alpha)^{-1})$.}
\end{figure}

The resonant scattering of quasiparticles from the fluctuating
vortex leads to interesting effects already at the one-loop level.
The quasiparticle LDOS is plotted in the Figure~\ref{LDOStotal} as a
function of energy at gradually increasing distances from the vortex
trap center. The resonant peak height is proportional to $\alpha^2$
at the trap center, and disappears over a length-scale comparable to
the extent of vortex zero-point oscillations $(m_\Tr{v}
\omega_\Tr{v})^{-1/2}$. This peak naturally lies at a sub-gap energy
whenever the length-scale set by the superconducting gap (comparable
to the vortex core size) is smaller than the amplitude of vortex
quantum oscillations. Despite the simplicity of our model, the
calculated LDOS has the same qualitative features as the LDOS
measured in the STM experiments \cite{aprile,hoffman,fischer}.

An important feature of the LDOS is the absence of a zero-energy
peak appearing in earlier computations
\cite{Wang95,machida,Franz98}. This appears to be a consequence of
the smallness of the vortex core \cite{ogata}. Even in our model, if
the vortex core is larger than the spatial extent of vortex
zero-point quantum fluctuations, the zero-energy peak appears in the
quasiparticle LDOS \cite{LorenzLDOS}. Therefore, the STM experiments
hint that it may not be the detailed structure of the vortex core
that is crucial for shaping the quasiparticle spectra near the
vortices in the cuprates; a viable alternative are quantum
fluctuations of vortices.

\begin{figure}
\includegraphics[width=1.65in]{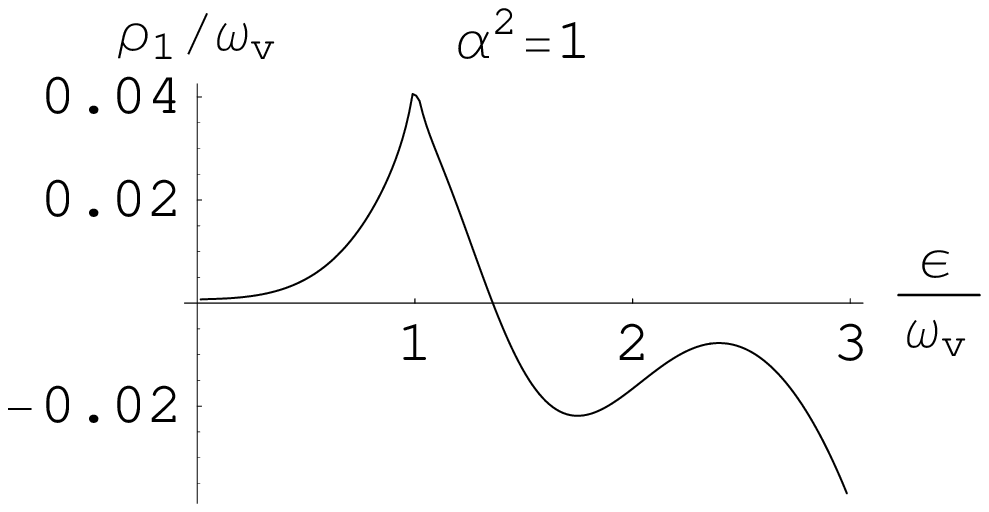}
\includegraphics[width=1.65in]{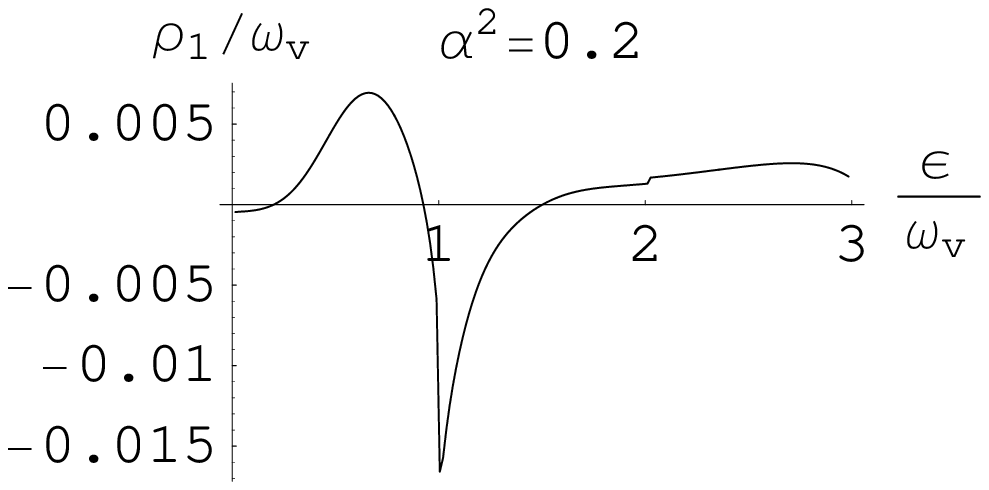}
\includegraphics[width=1.65in]{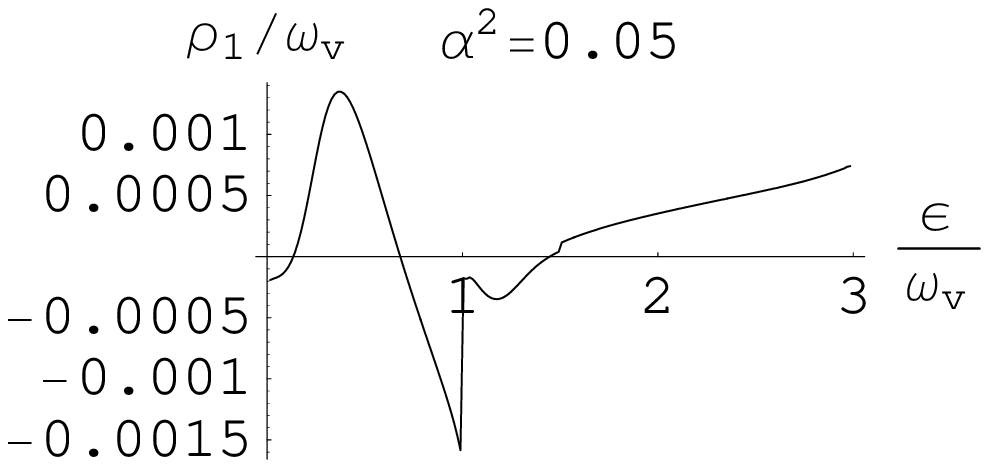}
\includegraphics[width=1.65in]{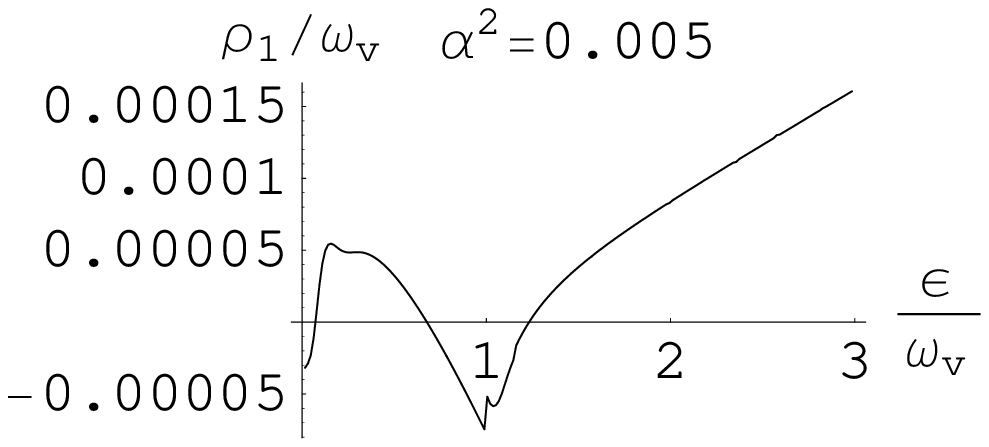}
\caption{\label{LDOS1}The one-loop correction $\rho_1$ to LDOS at
the vortex center as a function of energy. These plots show
evolution of $\rho_1(\epsilon)$ as the small parameter $\alpha^2$
changes. The magnitude of $\rho_1$ scales as $\alpha^2$.}
\end{figure}

The one-loop correction $\rho_1(\epsilon,r)$ of the LDOS
is calculated from the Fock self-energy:
\begin{equation}
    \begin{array}{ccc}
        \begin{minipage}{1.2in}
          \includegraphics[width=1.2in]{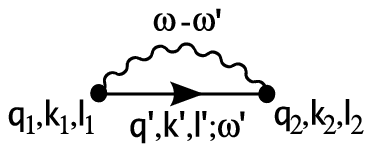}
        \end{minipage} &
        \ &
        \rho_1(\epsilon,r) = \frac{\omega_\Tr{v}}{\hbar v_\Tr{f}^2}
           \alpha^2 F_1 \left( \frac{\epsilon}{\hbar \omega_\Tr{v}},
                        \frac{\epsilon r}{\hbar v_\Tr{f}} ; \alpha \right)
    \end{array} \nonumber
\end{equation}
where the wiggly line represents a vortex excitation, and the
straight line a quasiparticle. $\rho_1(\epsilon,r)$ is
shown in the Figure~\ref{LDOS1} in order to emphasize all of
its features. In general, there is a discontinuity of LDOS at the
energy $\epsilon = \omega_\Tr{v}$. A major peak appears at the
energy $\epsilon \propto \alpha \omega_\Tr{v}$ when it does not
interfere with the discontinuity. The additional secondary features
of $\rho_1(\epsilon,r)$ are too small to be seen in the full
quasiparticle LDOS. The main peak moves toward lower energies
when the amplitude of vortex quantum fluctuations $\propto \alpha$
grows, indicating a short-lived resonant bound state between the
vortex and a quasiparticle.

Future even more sensitive experiments may provide more ways to
detect consequences of vortex quantum motion in $d$-wave
superconductors. The spatial extent of the LDOS modulations allows
measuring the energy scale $(\hbar \omega_\Tr{v} m_\Tr{v}
v_\Tr{f}^2)^{1/2}$, while resolving the thermally blurred
discontinuities of the LDOS would allow direct measurement of $\hbar
\omega_\Tr{v}$. Combined knowledge of both energy scales reveals the
vortex mass $m_\Tr{v}$ and the strength of the vortex trapping force
(inter-vortex interactions), which in turn can be related to other
parameters, such as the superconducting gap and magnetic field.
Effects of the moderately strong Magnus force merely reduce to
quantitative modifications of these energy scales.

In closing, we reiterate that the above model can also explain
\cite{CompOrd1} periodic LDOS modulations observed in recent STM
experiments \cite{hoffman,fischer}. Such modulations appear in the
LDOS over the region of vortex motion, and consequently the
experiments lead to an estimate of $m_\Tr{v}$ \cite{VorMass}.

\section{Conclusions}

Contribution of the nodal quasiparticles to the vortex mass is of
the order of an electron mass. Vortex friction arises only at finite
temperatures, or in presence of perturbations that create a finite
density of states at zero energy, such as disorder. Similarly,
quasiparticles whose density of states vanishes at zero energy do
not contribute any transversal forces on the vortex. Smallness of
vortex friction and mass at low temperatures have important
implications for the flux-flow in the ``normal state'' of $d$-wave
superconductors.

The quasiparticle LDOS in the vicinity of a localized quantum vortex
shows remarkable similarity to the experimental STM observations.
Absence of a zero-energy LDOS peak is an indication of the small
vortex core. Small sub-gap peaks appear in the LDOS as a result of
resonant scattering of quasiparticles from the quantum vortex. Some
secondary weak features also appear in the LDOS, including
discontinuities at energies that reflect the discrete spectrum of
the localized vortex. The future STM experiments might be sensitive
enough to detect these features, and allow measurements of the
vortex mass and vortex trapping forces.

\section{Acknowledgments}

We are very grateful to Zlatko Te\v sanovi\'c for several
stimulating discussions, and for freely sharing the results of
Ref.~\cite{Tesh2} prior to publication. We acknowledge useful
discussions with E.~Demler, B.~I.~Halperin, and A.~Kolezhuk. This
research was supported by NSF Grant DMR-0537077.

\end{document}